\documentclass[rnote]{aa} 
\usepackage{graphicx}
\usepackage{txfonts}
\begin{document}
   \title{Characterisation of the CAFOS linear 
   spectro-polarimeter\thanks{Based on observations 
   collected at the German-Spanish  Astronomical Center, 
   Calar Alto, jointly operated by the Max-Planck-Institut f\"ur 
   Astronomie Heidelberg and the Instituto de Astrofisica de Andalucia (CSIC).}}
  \subtitle{}
   \author{F. Patat\inst{1}
   \and    
    S. Taubenberger\inst{2}
}


   \institute{European Organization for Astronomical Research in the 
	Southern Hemisphere (ESO), Karl-Schwarzschild-Str.  2,
              85748, Garching b. M\"unchen, Germany
              \email{fpatat@eso.org}
              \and 
              Max-Planck-Institut f\"ur Astrophysik, Karl-Schwarzschild-Str. 1,
              85741, Garching b. M\"unchen, Germany
}

   \date{Received December, 2010; accepted ...}

\abstract{}{}{}{}{} 
 
  \abstract
   {}
   {This research note presents a full
   analysis of the CAFOS polarimeter mounted at the Calar Alto 2.2m
   telescope. It also provides future users of this mode with all
   necessary information to properly correct for instrumental effects
   in polarization data obtained with this instrument.}
  {The standard stars BD+59d389 (polarized) and
   HD~14069 (unpolarized) were observed with CAFOS in November, 2010, using
   16 half-wave plate angles. The linear spectropolarimetric
   properties of CAFOS were studied using a Fourier Analysis of 
   the resulting data. }
   {CAFOS shows a roughly constant instrumental polarization  at the level
   of $\sim$0.3\% between 4000 and 8600 \AA. Below 4000 \AA\/ the spurious 
   polarization grows to reach $\sim$0.7\% at 3600 \AA. This instrumental 
   effect is most likely produced by the telescope optics, and appears to be
   additive.  The Wollaston prism presents a clear deviation from the
   ideal behavior. The problem is largely removed by the usage of at least 4
   retarder plate angles. The chromatism of the half-wave plate causes
   a peak-to-peak oscillation of $\sim$11 degrees in the polarization
   angle. This can be effectively corrected using the tabulated values
   presented in this paper. The Fourier analysis shows that the $k\neq$0,4
   harmonics are practically negligible between 3800 and 7400 \AA.}
   {After correcting for instrumental polarization and retarder plate
    chromatism, with 4 half-wave plate angles CAFOS can reach an
    rms linear polarization accuracy of about 0.1\%.}

   \keywords{Techniques: polarimetry - Instrumentation: polarimeters}

\authorrunning{F. Patat and S. Taubenberger}
\titlerunning{Characterisation of CAFOS spectro-polarimeter}

   \maketitle
%

\section{\label{sec:intro}Introduction}

In the course of the observational campaign on the bright Supernova
2010jl we obtained spectropolarimetry of this object using the Calar
Alto Faint Object Spectrograph (CAFOS), mounted at the 2.2 m telescope
in Calar Alto, Spain (Meisenheimer \cite{cafos}). The results were
published in Patat et al. (\cite{patat10}). The polarimetric mode of
CAFOS has not been used very extensively, and mostly in imaging mode
(see Greiner et al. \cite{greiner} for an example). As we could not
find a proper characterisation of the instrumental effects in the
literature, during the campaign on SN~2010jl we ran a full analysis of
the instrument. This is presented here with the aim of making it
available to a wider community, who might find it useful for future
spectropolarimetric observations with this instrument.

Dual-beam polarimeters like CAFOS are composed by a half-wave retarder
plate (HWP) followed by the analyzer, which is a Wollaston prism (WP)
producing two beams with orthogonal directions of polarization,
usually indicated as ordinary (O) and extraordinary (E) beams. With
this instrumental setup, the Stokes parameters $Q$ and $U$ are derived
measuring the intensities in the O and E beams ($f_{O,i}$, $f_{E,i}$)
at a given set of HWP angles $\theta_i$ (for a general overview see
Patat \& Romaniello \cite{patat06}, and references therein).

This is typically achieved through the normalized flux differences
$F_i$,

\begin{displaymath}
F_i=\frac{f_{O,i}-f_{E,i}}{f_{O,i}+f_{E,i}}.
\end{displaymath}

For an ideal polarimeter, the normalized flux differences obey to the
following relation: $F_i=P \cos(4\theta_i - 2\chi)$, where
$P=\sqrt{Q^2+U^2}$ is the polarization degree, and
$\chi=\frac{1}{2}\arctan(U/Q)$ is the polarization position
angle. Although any set of angles $\theta_i$ is in principle suitable
for obtaining $Q$ and $U$, the optimal choice is
$\theta_i=\frac{\pi}{8} i$. In these conditions one has that:

\begin{displaymath}
Q=\frac{2}{N}\sum_{i=0}^{N-1} F_i \cos \left (  \frac{\pi}{2} i\right ) \\
\end{displaymath}
\begin{displaymath}
U=\frac{2}{N}\sum_{i=0}^{N-1} F_i \sin \left (  \frac{\pi}{2} i\right ),
\end{displaymath}

\noindent where $N$ is the number of HWP angles. As $F_i$ can be
thought as a co-sinusoidal signal modulated by the rotation of the HWP
with a fundamental period 2$\pi$, these can be rewritten as a Fourier
series (see Fendt et al. \cite{fendt}):

\begin{equation}
F_i = a_0 +
\sum_{k=1}^{N/2} \left [ a_k \cos \left( k \frac{2\pi i}{N}\right) +
b_k \sin \left( k \frac{2\pi i}{N}\right) \right],
\end{equation}

\noindent where $a_k$ and $b_k$ are the Fourier coefficients. The
Fourier analysis is particularly useful when $N$=16; under these
circumstances, the polarization signal is carried by the $k$=4
component. In an ideal system, all other components are rigorously
zero. Therefore, non-null Fourier coefficients for $k\neq$4 signal
possible problems in the polarimeter. For the meaning of the various
components the reader is referred to Fendt et al. (\cite{fendt}).

\begin{figure}
\centerline{ \includegraphics[width=90mm]{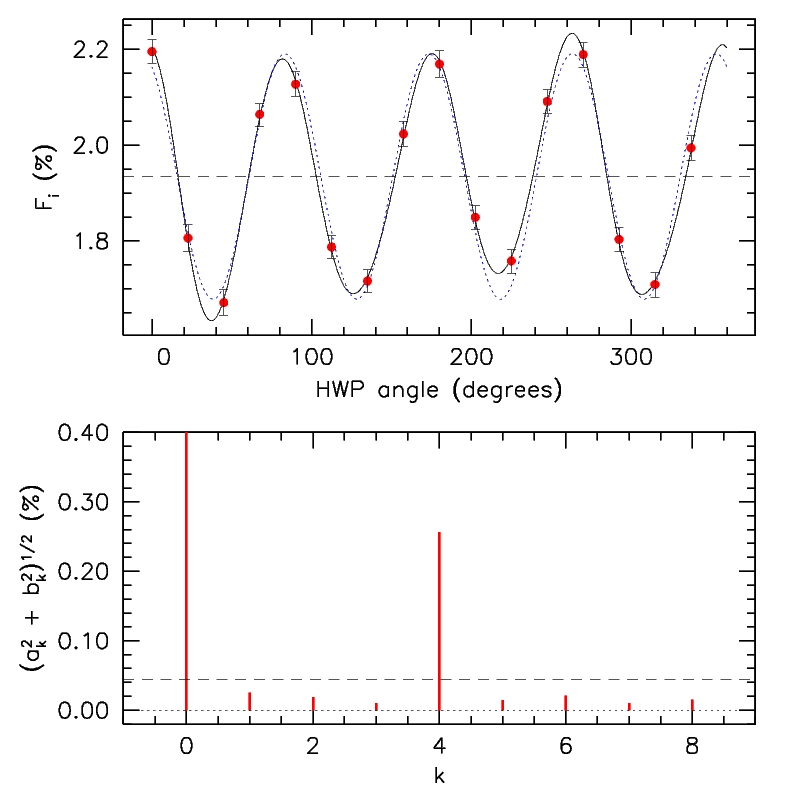} }
\caption{\label{fig:unpol}Fourier analysis applied to the unpolarized
  standard star HD~14069 at 5500 \AA\/ (200 \AA\/ bin). {\bf Top:} normalized
  flux differences. The curves trace the partial reconstruction using
  8 harmonics (solid) and the fourth harmonic only (dotted). The
  dashed horizontal line is placed at the average of the $F$ values
  ($a_0$). {\bf Bottom:} harmonics power spectrum. The dashed line indicates
  the 5-sigma level of the uncertainty.}
\end{figure}

\section{\label{sec:obs}Observations and data reduction}

The observations were carried out with CAFOS (Meisenheimer
\cite{cafos}). In this multi-mode instrument, equipped with a
2K$\times$2K SITe-1d CCD (24$\mu m$ pixels, 0.53 arcsec/pixel),
polarimetry is performed by introducing into the optical path a WP
(18$^{\prime\prime}$ throw) and a super-achromatic HWP, between the
collimator and the grism. For our study we observed the polarized star
BD+59d389 ($P(V)$=6.70$\pm$0.02\%, $\chi$=98.1 degrees; Schmidt et
al. \cite{schmidt}), and the unpolarized star HD~14069
($P(V)$=0.02$\pm$0.02\%; Schmidt et al. \cite{schmidt}) on 2010,
November 18.8 UT.  All spectra were obtained with the low-resolution
B200 grism coupled with a 1.0 arcsec slit, giving a spectral range
3300-8900 \AA, a dispersion of $\sim$4.7 \AA\/ px$^{-1}$, and a FWHM
resolution of 14.0 \AA. The slit was aligned along the N-S direction.
To enable the Fourier analysis up to the 8-th harmonic we used $N$=16
half-wave plate angles (0, 22.5, ..., 337.5).  The exposure times were
180 seconds per HWP angle for both standard stars.

Data were bias and flat-field corrected, and wavelength calibrated
using standard tasks within IRAF\footnote{IRAF is distributed by the
  National Optical Astronomy Observatories, which are operated by the
  Association of Universities for Research in Astronomy, under
  contract with the National Science Foundation.}. The Fourier
analysis was carried out using specific routines written by us.

\section{\label{sec:inspol}Instrumental polarization}

\begin{figure}
\centerline{ \includegraphics[width=90mm]{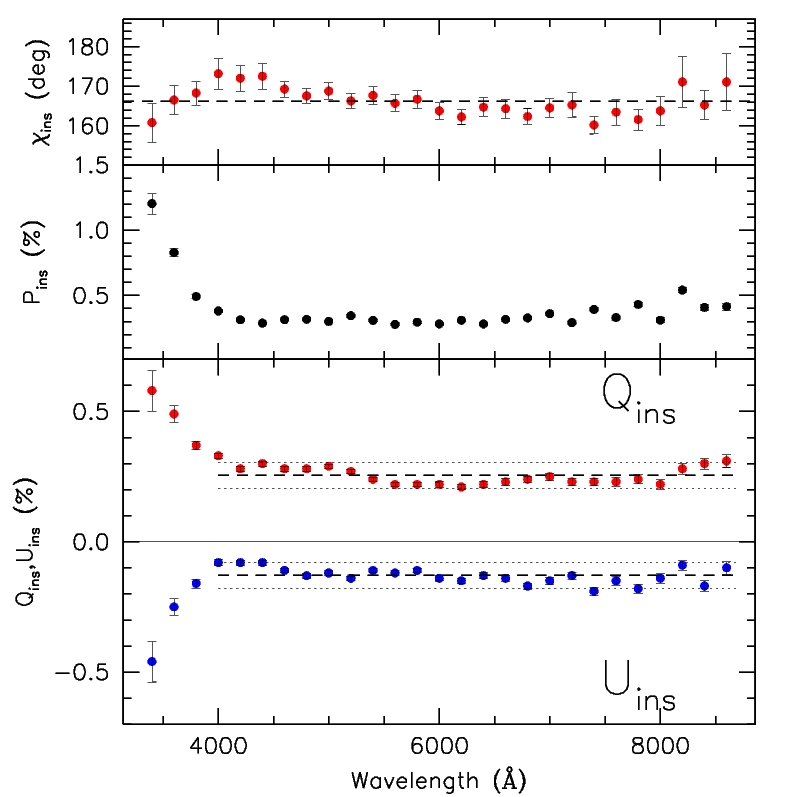} }
\caption{\label{fig:inspol} CAFOS instrumental polarization. {\bf Upper
  panel}: instrumental polarization position angle. The dashed line
  indicates the average value. {\bf Mid panel}: instrumental polarization
  degree. {\bf Lower panel}: instrumental Stokes parameters. The dashed
  lines indicate the average value of $Q$ and $U$ in the wavelength
  range 4000--8600 \AA, while the dotted lines mark the $\pm$0.05\%
  deviations from the average value.}
\end{figure}

To characterize the instrumental polarization of CAFOS, we first
analyzed the data obtained for the unpolarized standard star. The
result of the Fourier analysis is presented in Fig.~\ref{fig:unpol}
for a 200 \AA\/ wide bin centered at 5500 \AA. The normalized flux
differences show a marked modulation (upper panel), well reproduced by
a sinusoidal function. The power spectrum (lower panel) displays a
neat peak at the $k$=4 overtone, corresponding to a linear
polarization signal (see Sec.~\ref{sec:intro}), reaching $P$=0.26\%
(the $k$=0 term is also non null, but we will discuss this in
Sect.~\ref{sec:fourier}). The fact that the signal is modulated by the
retarder plate rotation implies that the source of instrumental
polarization precedes the HWP along the optical path. Therefore, most
likely the observed polarization arises within the collimator and/or
the telescope mirrors. For instance, inhomogeneities in the mirror
coatings can break circular symmetry, and lead to an incomplete
cancellation of the linear polarization generated by reflections (see
Tinbergen \cite{tinbergen} and Leroy \cite{leroy} for general
introductions to the subject). Such a system would behave as a partial
polarizer, characterized by a certain position angle ($\chi_{ins}$)
that does not depend on wavelength, but only on the geometry of the
system asymmetry. In general, the effect of the instrumental
polarization depends on the Stokes vector that characterizes the input
signal, and this makes the correction for instrumental polarization
particularly difficult. However, when the instrumental polarization is
much smaller than 1, the effect is additive, and the spurious signal
can be removed subtracting it vectorially from the measured one (see
for instance Patat \& Romaniello \cite{patat06} for the case of
VLT-FORS1).

\begin{figure}
\centerline{ \includegraphics[width=90mm]{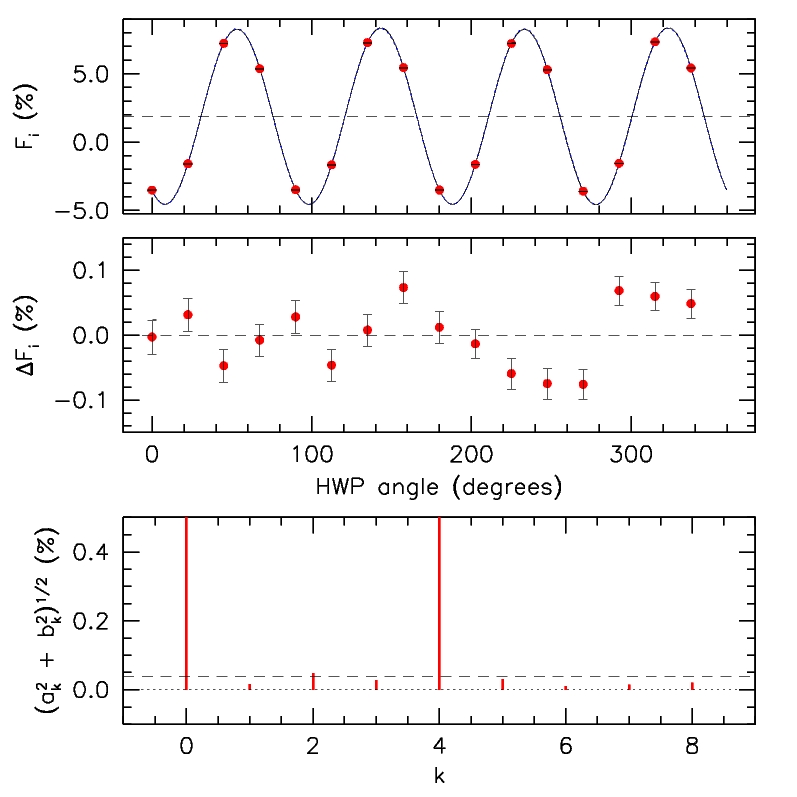} }
\caption{\label{fig:polstd}Fourier analysis applied to the polarized
  standard star BD+59d389 at 5500 \AA\/ (200 \AA\/ bin). {\bf Top:} normalized
  flux differences. The curves trace the partial reconstruction using
  8 harmonics (solid) and the fourth harmonic only (dotted). The
  dashed horizontal line is placed at the average of the $F$ values
  ($a_0$). {\bf Middle:} residuals from the reconstruction using the $k$=4
  harmonic.  {\bf Bottom:} harmonics power spectrum. The dashed line
  indicates the 5-sigma level uncertainty.}
\end{figure}

The presence of a constant position angle is confirmed by the Fourier
analysis run across the whole wavelength range covered by our
observations. In Fig.~\ref{fig:inspol} we present the values of
$Q_{ins}$ and $U_{ins}$ derived within 200 \AA\/ wide bins between
3400 and 8600 \AA\/ (lower panel), and the implied position angle
$\chi_{ins}$ (upper panel). The average value of $\chi_{ins}$ is 166.3
degrees, and the RMS deviation of the single measurements is 3.6
degrees. The smooth oscillation seen in the position angle is related
to the chromatism of the HWP retardance (see Sect.~\ref{sec:hwp}).
As far as the polarization is concerned, this reaches 0.74$\pm$0.08\%
at 3400 \AA, and it rapidly decreases to 0.33$\pm$0.01\% at 4000 \AA,
to remain constant to within 0.05\% up to 8600 \AA. The reason for the
marked increase seen bluewards of 4000 \AA\/ is not clear, but it
might be related to the decrease of efficiency in the anti-reflexion
coatings of the collimator lenses.

The average values of the instrumental Stokes parameters above 4000
\AA\/ are $\langle Q_{ins}\rangle$=+0.25$\pm$0.03\%, and $\langle
U_{ins}\rangle$=$-$0.13$\pm$0.03\% respectively, leading to an average
polarization $P_{ins}$=0.28$\pm$0.03\%. The wavelength range below
3800 \AA\/ is affected by other instrumental problems which make it
hardly usable with the typical set of 4 HWP angles (see next section).
Therefore, this constant correction is sufficient to guarantee the
removal of the instrumental polarization with a maximum error of
0.05\%, which is comparable to the maximum accuracy one can reach with
CAFOS with 4 HWP angles (see next section).

We remark that the instrumental polarization correction derived here
is strictly valid only for an object placed on the CAFOS reference
pixel used for the acquisition onto the 1.0 arcsec slit. With the
present analysis we cannot exclude position-dependent effects,
similarly to what happens in the FORS instruments (Patat \& Romaniello
\cite{patat06}).

\begin{figure}
  \centerline{ \includegraphics[width=90mm]{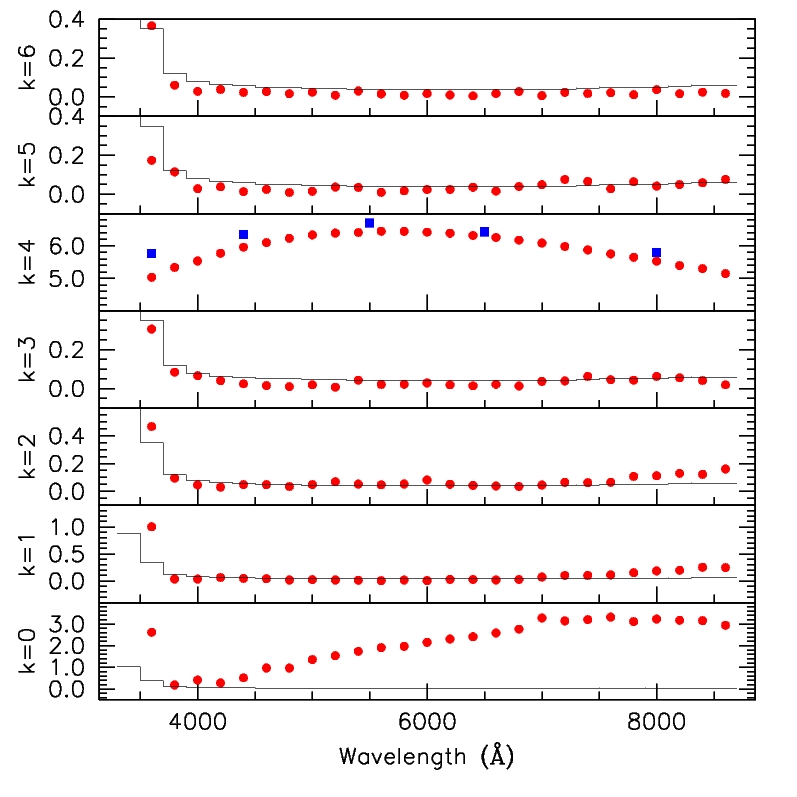} }
\caption{\label{fig:fullanal}Power spectrum of the first 6 harmonics
  as a function of wavelength (ordinate scale is in \%). The solid
  thin lines trace the 5-$\sigma$ confidence level. The filled squares
  in the $k$=4 plot are the broad-band polarization measurements by
  Schmidt et al. (\cite{schmidt}).}
\end{figure}

\section{\label{sec:fourier}Fourier analysis}

For the Fourier analysis of the CAFOS polarimetric performances we
have used the data obtained for the polarized standard
star. Fig.~\ref{fig:polstd} shows an example for a 200 \AA\/ wide bin
centered at 5500 \AA. The only components which show a statistically
significant power are $k$=0 and $k$=4; there is a hint of a nun-null
$k$=2 component, which is related to the so-called pleochroism (Fendt
et al. \cite{fendt}; Patat \& Romaniello \cite{patat06}), but this is
only marginally significant at the 5-$\sigma$ level. The original
signal can be reconstructed using only the $k$=4 harmonic, with
maximum residuals $\Delta F_i$ of $\sim$0.1\%. This implies that 4 HWP
angles are sufficient to the derive the Stokes parameters with a
maximum error of this order. The polarization degree derived using 16
HWP angles at 5500 \AA\/ is 6.43$\pm$0.01\%. After applying the
instrumental polarization correction described in the previous section
this value becomes 6.6$\pm$0.1\%. This is fully consistent with
  the reference value 6.70$\pm$0.02\% measured in the V passband
  (Schmidt et al. \cite{schmidt}).

In the example illustrated in Fig.~\ref{fig:polstd} we find
$a_0$=1.88$\pm$0.01\% (the corresponding value derived from the
  unpolarized standard is 1.93$\pm$0.01\%; see also
Fig.~\ref{fig:unpol}, upper panel). This indicates that the WP
deviates from the ideal case, in that an unpolarized incoming beam is
not exactly split into two identical fractions (see Patat \&
Romaniello \cite{patat06}, their Sect.~7). As a consequence, using
only 2 HWP angles (which is the minimum set needed to fully
reconstruct the Stokes vector) would lead to a very significant error
on the final result.

\begin{figure}
  \centerline{ \includegraphics[width=90mm]{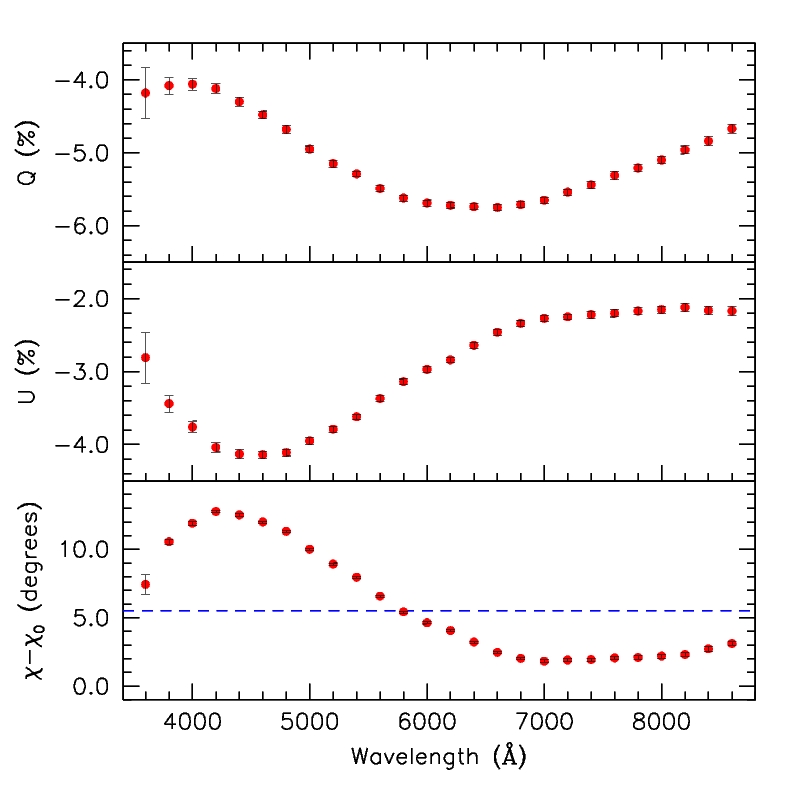} }
\caption{\label{fig:chromat} Instrumental polarization corrected
  Stokes parameters $Q$ (top panel) and $U$ (mid panel) for
  BD+59d389. The bottom panel shows the phase retardance variation as
  a function of wavelength.}
\end{figure}

To study the instrumental performance as a function of wavelength, we
have run the same analysis within 200 \AA\/ wide bins between 3400 and
8600 \AA. The result for the first 6 harmonics is shown in
Fig.~\ref{fig:fullanal}. The $k$=0 component is always significant,
exceeding $\sim$3\% at 7500 \AA, but this is fairly well corrected if
the data set includes at least 4 HWP positions. As for components
$k$=1 and 2, these are detected at a significant level below 3800
\AA\/ and above 7000 \AA. At 3600 \AA\/ the usage of 4 HWP angles
leads to errors larger than 0.3\%, making data bluewards of 3800 \AA\/
hardly usable. At the red edge, deviations are below 0.2\% bluewards of
7400 \AA, while they can exceed 0.3\% above 8200 \AA. 

As the $k$=4 component of the power spectrum is the linear
polarization degree, its wavelength dependence can be directly
compared to the broad band values available in the literature (Schmidt
et al. \cite{schmidt}). These are overplotted in the $k$=4 panel of
Fig.~\ref{fig:fullanal} (filled squares). As expected based on the
estimates of the instrumental polarization (see
Sect.~\ref{sec:inspol}), there is a difference of about 0.3\% above
4000 \AA. The value corresponding to the $U$ passband shows a larger
deviation (0.7\%), which is consistent with the increase of the
instrumental polarization seen below 4000 \AA\/ (see
Fig.~\ref{fig:inspol}). It is worth noting that, as the
  polarization signals of the star and the instrument are close to
  orthogonal, the corrected value is higher than the measured one.

\section{\label{sec:hwp}HWP chromatism}

Although the retarder plate deployed in CAFOS is super-achromatic, the
phase retardance is expected to deviate from an ideal behavior as a
function of wavelength. To quantify this effect we have used the
polarized standard as reference. For this star the polarization
position angle is constant to within 0.1 degrees in the UBVRI domain,
the average value being $\chi_0$=98.2$\pm$0.1 degrees (Schmidt et
al. \cite{schmidt}). Therefore, if $Q_{obs}$ and $U_{obs}$ are the
measured Stokes parameters, the phase retardance variation across the
wavelength range can be computed as $\Delta \chi=\chi-\chi_0$, where
$\chi=\frac{1}{2}\arctan [(U_{obs}-U_{ins})/(Q_{obs}-Q_{ins})]$. The
result is plotted in Fig.~\ref{fig:chromat}, and the values listed in
Table~\ref{tab:chromat}.

\begin{table}
\tabcolsep 3 mm
\begin{tabular}{ccccccc}
\hline
$\lambda$ & $\Delta \chi$ & $\sigma$ & & $\lambda$ & $\Delta \chi$ & $\sigma$\\
(\AA) & (deg) & (deg) & & (\AA) & (deg) & (deg)\\
\hline
3600 &    7.44 &  0.71 & & 6200 &    4.06 &  0.08 \\
3800 &   10.55 &  0.20 & & 6400 &    3.22 &  0.09\\
4000 &   11.90 &  0.12 & & 6600 &    2.46 &  0.09\\
4200 &   12.76 &  0.09 & & 6800 &    2.02 &  0.10\\
4400 &   12.51 &  0.07 & & 7000 &    1.82 &  0.10\\
4600 &   12.00 &  0.07 & & 7200 &    1.91 &  0.10\\
4800 &   11.31 &  0.07 & & 7400 &    1.93 &  0.12\\
5000 &   10.00 &  0.07 & & 7600 &    2.06 &  0.13\\
5200 &    8.91 &  0.07 & & 7800 &    2.09 &  0.13\\
5400 &    7.95 &  0.06 & & 8000 &    2.19 &  0.13\\
5600 &    6.57 &  0.07 & & 8200 &    2.30 &  0.15\\
5800 &    5.42 &  0.07 & & 8400 &    2.72 &  0.16\\
6000 &    4.63 &  0.08 & & 8600 &    3.11 &  0.16\\
\hline
\end{tabular}
\caption{\label{tab:chromat}HWP retardance variation as a function of
  wavelength.}
\end{table}

Having these values at hand, the corrected Stokes parameters $Q_c$ and
$U_c$ can be obtained by the following rotation:

\begin{displaymath}
Q_c = Q \cos 2\Delta \chi + U \sin 2\Delta \chi
\end{displaymath}

\begin{displaymath}
U_c = U \cos 2\Delta \chi - Q \sin 2\Delta \chi ,
\end{displaymath}

\noindent where $Q$ and $U$ are the instrumental polarization
corrected Stokes parameters. Alternatively, the position angle
obtained from $Q$ and $U$ can be corrected subtracting $\Delta \chi$.

Usually the zero-point of the HWP angle is set so that $\theta$=0
corresponds to a null astronomical position angle in the plane of the
sky around the central wavelength. This is not the case in CAFOS, as
the deviation at 6000 \AA\/ is about 5.5 degrees, and it is never zero
between 3600 and 8600 \AA\/ (Fig.~\ref{fig:chromat}, bottom panel).
However, given the way we have computed $\Delta \chi$, this correction
will give position angles in the plane of the sky, with $\chi$=0
corresponding to the N-S direction.

\section{\label{sec:conc}Conclusions}

In this note we presented a full analysis of the linear polarization
properties of CAFOS. Although the instrument appears to suffer from a
significant spurious polarization, this can be removed to within
$\sim$0.1\%. The effect appears to be additive, and can be therefore
easily corrected by vectorially subtracting the instrumental
component on the Stokes $Q,U$ plane.

As is typical of other dual-beam polarimeters (see for instance the
case of FORS1, Patat \& Romaniello \cite{patat06}), the Wollaston
prism departs from the ideal case. In the worst case the fraction of
light in the ordinary and extraordinary beams for an unpolarized
incoming signal deviates by $\sim$2\% from the theoretical 50/50
ratio. However, this defect is largely removed by the adoption of 4
retarder plate angles during the observations. Using the minimum set
(2 HWP angles) leads to large errors, especially in the case of low
polarizations ($\sim$1\%), and it is therefore strongly discouraged.

The Fourier analysis shows that all harmonics with $k\neq$0,4 are
negligible in the wavelength range 3800--7400 \AA, where a rms accuracy
of 0.1\% can be reached with a sufficient signal-to-noise. This can be
considered as the instrumental limit attainable with CAFOS with 4 HWP
angles, and within this spectral range. Below 3800 \AA\/ the
polarimetric properties rapidly degrade, requiring a larger number of
HWP angles. The same applies, though to a smaller extent, to the
region redwards of 8200 \AA.

For this work we have used data obtained with the B200 grism. Because
of its tilted surfaces, the grism can act as a poor linear
polarizer. Since in CAFOS this is placed after the analyzer, the
spurious polarization produced by transmission is not modulated by the
HWP rotation, and hence the redundancy in the retarder-plate position
effectively removes it (see Patat \& Romaniello \cite{patat06}). The
exact effect produced by the grism depends on its properties. However,
the conclusions reached in this paper do not depend on the grism,
provided that the data are obtained using at least 4 HWP angles.

In general, CAFOS appears to be perfectly suitable for linear
polarization studies aiming at accuracies of a few 0.1\%, making it a
valid instrument for bright objects. As a term of reference, an
accuracy of $\sim$0.1\% per resolution element ($\sim$50 \AA) was
reached for SN~2010jl ($V\sim$13.5), with 4 exposures of 40 minutes
each (Patat et al. \cite{patat10}).

\begin{acknowledgements}
We wish to thank the personnel at Calar Alto Observatory for obtaining
the extra calibrations required for the full polarimetric
characterisation of CAFOS. This work has been done in the framework of
the European Large Collaboration {\it Supernova Variety and
Nucleosynthesis Yields} (http://graspa.oapd.inaf.it/),
whose members are acknowledged.
 
\end{acknowledgements}

\end{document}